\documentclass[shortbibliography,twocolumn,prl,aps,superscriptaddress,showpacs,amsmath,amssymb,floatfix]{revtex4-2}
\usepackage{graphicx}
\usepackage{amssymb}
\usepackage{amsmath}
\usepackage{epsfig}
\usepackage{color}
\usepackage{mathtools}
\usepackage[colorlinks=true, letterpaper=true, pdfstartview=FitV, linkcolor=blue, citecolor=blue, urlcolor=blue]{hyperref}
\usepackage{physics}
\usepackage{siunitx}
\usepackage{bm}
\usepackage{multirow}
\usepackage[dvipsnames]{xcolor}
\definecolor{mypurple}{RGB}{112, 48, 160}
\definecolor{myblue}{RGB}{0, 112, 192}
\usepackage{booktabs}
\begin{document}

\title{Proper Definition of Intrinsic Nonlinear Current}

\author{Cong Xiao}
\email{congxiao@fudan.edu.cn}
\thanks{These two authors contributed equally}
\affiliation{Interdisciplinary Center for Theoretical Physics and Information Sciences (ICTPIS), Fudan University, Shanghai 200433, China}

\author{Jin Cao}
\thanks{These two authors contributed equally}
\affiliation{Research Laboratory for Quantum Materials, IAPME, University of Macau, Taipa, Macau, China}

\author{Qian Niu}
\affiliation{International Centre for Quantum Design of Functional Materials,
CAS Key Laboratory of Strongly-Coupled Quantum Matter Physics, and Department of Physics,
University of Science and Technology of China, Hefei, Anhui 230026, China}

\author{Shengyuan A. Yang}
\affiliation{Research Laboratory for Quantum Materials, IAPME, University of Macau, Taipa, Macau, China}

\begin{abstract}
We show that the three commonly employed approaches that define the same dc (or low-frequency) intrinsic linear anomalous Hall response actually lead to different results for intrinsic nonlinear transport. The difference is due to an intrinsic anomalous distribution (IAD). It originates from a nonlinear field effect during scattering, but its value is completely independent of scattering, because it represents the local equilibration of electron wave packets with field corrected energy. The proper definition of intrinsic current that is detectable in experiment must incorporate the effect of IAD. We also show that IAD is indispensable for consistency with fundamental physical relations.
In addition, we predict that under ac driving, the intrinsic reponses in rectified and double-frequency channels exhibit distinct frequency dependence, for which we estimate the signals that can be probed in antiferromagnetic CuMnAs.

\end{abstract}

\maketitle

Intrinsic responses, which depend only on band structures of materials, are of crucial importance in condensed matter physics, since they can be evaluated accurately with first-principles techniques, offer benchmarks for experiments, and probe interesting geometric band properties. A prominent example is anomalous Hall effect \cite{nagaosa2010,sinitsyn_semiclassical_2007}, whose intrinsic contribution manifests the Berry curvature of band structures \cite{jungwirth2002,nagaosa2002}. In that context, there have been three different approaches to define what is  intrinsic response in dc diffusive transport. (D1) The first approach is to consider the naive ``clean limit'', without any scattering. For example, in the semiclassical framework~\cite{xiao2010,sinitsyn_semiclassical_2007}, this means one directly takes the equilibrium distribution
$f_0$ in the calculation of response. (D2) The second is to define intrinsic response as the extrapolation of the ac Hall response from the high-frequency regime ($\omega\gg \tau^{-1}$, where $\tau$ is the scattering time) to the dc limit, with $\tau^{-1}\rightarrow 0$ faster than $\omega\rightarrow 0$ \cite{nagaosa2010}. (D3) The third one is an operational definition, namely, one fully considers scattering and defines intrinsic response as the scattering independent component of the total response. For linear anomalous Hall transport, the three approaches yield the same result hence are considered equivalent.

Recently, there has been tremendous interest in nonlinear Hall effect, for which the charge current response $j\propto E^2$ is quadratic in the driving $E$ field \cite{gao2014,fu2015,facio2018,du2018,ma2019,kang2019,du2019,Sodemann2019,Shao2020,Isobe2020,He2021quantum,kumar2021room,Zhang2021,Tiwari2021,Lu2021,ortix2021nonlinear,watanabe2021,Sodemann2021,wang2021,liu2021,He2022graphene,Liu2022PRL,sinha2022,Liao2023,Gao2023QM,Wang2023QM,Han2024room,Huang2023,Amit2023,Kang2023switchable,Song2023ASK,Su2024,Yan2024,Fang2024,Jia2024,Lee2024,Ortix2024,Liu2024}. Under broken time reversal symmetry, this effect also allows an intrinsic response \cite{gao2014}.
{\color{black}Despite intensive studies, there remains controversy on the very definition of this intrinsic response, which
has caused conflicting interpretations of experimental results. For example, two pioneering experiments Ref.~\cite{Gao2023QM} and Ref.~\cite{Wang2023QM}, both on even-layer MnBi$_2$Te$_4$, are interpreted with two differently defined intrinsic nonlinear currents [corresponding to (D3) and (D1), respectively, as we see below] which even possess distinct symmetry characters.
This issue of proper definition of intrinsic nonlinear current has generated lasting confusion and become a central problem of nonlinear transport,
which requires an urgent solution.
}

%
%
%
%
%

\begin{table*}[ht]
\caption{Different results of intrinsic nonlinear current obtained from the three definitions. The second, third, and fourth columns correspond to (D1), (D2), and (D3), respectively. }
\label{Difference}%
\begin{tabular}{cccc}
\hline\hline
Component & without counting scattering \ & $\ \omega \tau \gg 1$ and $\omega \rightarrow 0$ \ & \ with scattering, expressions independent of scattering  \\ \hline
dc response & \ \ \ Eq. (\ref{jcl}) & \ $\frac{1}{2}$(Eq. (\ref{jcl})+Eq. (\ref{jin})) \ & \ \ \ Eq. (\ref{jin}) \\
ac rectified  & \ \ \ Eq. (\ref{jcl}) & \ \ $-$ & \ \ \ Eq. (\ref{jin}) \\
ac double-frequency  & \ \ \ Eq. (\ref{jcl}) & \ \ $-$ & \ \ \ $\omega  \ll 1/\tau$, Eq. (\ref{jin}); \ $\omega \gg 1/\tau$, Eq. (\ref{jcl}) \\ \hline\hline
\end{tabular}%
\end{table*}

In this work, we address this critical problem. We find that distinct from linear transport, the three approaches actually lead to different results for nonlinear transport (see Table~\ref{Difference}). This difference can be traced to
the field correction to the electron wave packet energy, which is quadratic in $E$.
Owing to this energy shift, in the presence of disorder scattering, electrons acquire a correction in distribution function, termed as the intrinsic anomalous distribution (IAD),
which turns out to be \emph{completely independent} of scattering. Physically, the corrected distribution which includes IAD can be naturally interpreted as the equilibrium state for an ensemble of electron wave packets with the corrected energy. It is crucial that the establishment of this new equilibrium requires scattering, but the resulting distribution itself has no dependence on scattering at all. It follows that for dc transport, (D3) fully captures IAD and gives a \emph{purely Hall} response;
(D1) must differ from (D3), since it does not consider any scattering effect; meanwhile, (D2) partially captures IAD and leads to a result sitting in-between. We argue that
(D3) is the proper definition of intrinsic nonlinear response, because IAD is an inseparable part of intrinsic response that is actually detected in dc or low-frequency experiment. In other words, (D3) is the appropriate one that should be used for interpreting experiments.

{\color{black}Furthermore, the notion of IAD is universal, and its significance is not limited to nonlinear transport.
We demonstrate that it is indispensable for any coherent theory which maintains consistency with fundamental physics relations and principles, such as the Einstein relation and the absence of dc current driven by a static uniform magnetic field.}

In addition, as a by-product of this new understanding, we predict that under ac driving,
the second-harmonic component of intrinsic nonlinear current may acquire different values in low- and high-frequency regimes. Using first-principles calculation, we estimate signals that can be detected in CuMnAs.


\textit{\color{blue}Preliminaries.} Consider diffusive transport under a dc driving $E$ field. The charge current density is (we set $e=\hbar=1$)
\begin{equation}\label{current}
  \bm j=-\sum_l f_{l} \bm v_{l},
\end{equation}
where $l\equiv (n,\bm k)$ is a composition of band and crystal momentum labels, and $\bm v_{l}$ is the velocity of electron wave packet. Up to second order in $E$, the scattering-independent part of velocity is given by \cite{gao2014,gao2019}
\begin{equation}\label{vv}
  \bm v_{l}=\partial_{\bm k}\tilde{\varepsilon}_l+\bm E\times \tilde{\bm \Omega}_{l}.
\end{equation}
Here,
\begin{equation}
  \tilde{\varepsilon}_l=\varepsilon_l+\delta \varepsilon_l
\end{equation}
is the field-corrected wave packet energy: Besides the unperturbed band energy $\varepsilon_l$, there is a second-order field correction \cite{liu2022,Xiao2022NLSOT,Jia2024}
\begin{equation}\label{energy shift}
  \delta \varepsilon_l=-\frac{1}{2}G_{ab}^l E_a E_b,
\end{equation}
where summation over repeated Cartesian indices $a$ and $b$ is implied, and $G_{ab}^l$ is known as Berry-connection polarizability (BCP) tensor \cite{liu2021,liu2022}. It is a gauge-invariant intrinsic band property, given by \cite{gao2014}
\begin{equation}\label{k-BCP}
  G_{ab}^l=2 \text{Re}\sum_{n'\neq n}\frac{V_a^{nn'}V_b^{n'n}}{(\varepsilon_{n}-\varepsilon_{n'})^3},
\end{equation}
where $V^{nn'}$ is the interband velocity matrix element, and quantities on both sides are evaluated at the same $k$ point. In Eq.~(\ref{vv}), the Berry curvature $\tilde{\bm \Omega}_{l}$ includes the first-order field correction, i.e., $\tilde{\bm \Omega}=\bm \Omega+\delta \bm \Omega$, where $\bm \Omega$ is for unperturbed bands, and the correction 
$\delta \Omega_a=\epsilon_{abc}\partial_{k_b}G_{cd}E_d$ ($\epsilon_{abc}$ is the totally antisymmetric tensor) is also determined by BCP \cite{gao2014,wang2021,liu2021}.

To obtain the intrinsic nonlinear current, in approach (D1), one simply put $f_l=f_0(\varepsilon_l)$, substitute Eq.~(\ref{vv}) into (\ref{current}), and collect terms of $E^2$ order. The result is
\begin{equation}\label{jcl}
  \bm j^{\mathrm{cl}}=-\sum_l f_0(\varepsilon_l)\Big(\partial_{\bm k} \delta \varepsilon_l+\bm E\times \delta \bm \Omega_l\Big).
\end{equation}

One may already spot an ambiguity in the treatment above, namely, since we are effectively dealing with a bunch of field-dressed wave packets with energies $\tilde{\varepsilon}_l$, it seems more appealing to use their equilibrium distribution $f_0(\tilde{\varepsilon}_l)$ in Eq.~(\ref{current}) for the intrinsic current. By doing so and performing the expansion $f_0(\tilde{\varepsilon})=f_0({\varepsilon})+f_0'({\varepsilon})\delta \varepsilon$, one obtains
a different and simpler intrinsic current~\cite{supp}:
\begin{equation}\label{jin}
  \bm j^{\mathrm{in}}=-\sum_l f_0(\varepsilon_l)(\bm E\times \delta \bm \Omega_l),
\end{equation}
which is a purely Hall (hence dissipationless) response.

The two currents differ by
\begin{equation}
  \bm j^{\mathrm{in}}-\bm j^{\mathrm{cl}}=-\sum_l g_l \partial_{\bm k}\varepsilon_l,
\end{equation}
where
\begin{equation}\label{iad}
  g_l=f_0(\tilde{\varepsilon}_l)-f_0({\varepsilon}_l)
\end{equation}
is the difference between the two distributions, which we refer to as IAD.

$\bm j^{\mathrm{in}}$ in Eq.~(\ref{jin}) has been obtained in several pioneering works on this effect \cite{gao2014,wang2021,liu2021}. We shall show that it corresponds to approach (D3) in dc and low-frequency transport. IAD is the key for the difference between $\bm j^{\mathrm{in}}$ [from (D3)] and $\bm j^{\mathrm{cl}}$ [from (D1)].
It must be emphasized that this difference
is dramatic: The two results even have distinct symmetry characters, e.g.,
$\bm j^{\mathrm{in}}$ is purely Hall and is forbidden by any rotational symmetry normal to the transport plane (such as  $C_{3z}$) \cite{Gao2023QM}, whereas $\bm j^{\mathrm{cl}}$ is not.


\textit{\color{blue}Intrinsic anomalous distribution.} Now, the question is: Where does the IAD $g_l$ come from?

The original system in the absence of driving field is in the equilibrium state described by $f_0(\varepsilon_l)$. When one gradually turns on the $E$ field, the electron wave packets are dressed, with their energies changed to $\tilde{\varepsilon}_l$.
Clearly, $f_0(\varepsilon_l)$ makes no longer an equilibrium state for particles with these renormalized energies, so it should relax towards an (new) equilibrium corresponding to $f_0(\tilde{\varepsilon}_l)$.
This is the part of the distribution function in addition to the non-equilibrium part due to the drift of Fermi surface by field and other extrinsic effects. Evidently, the relaxation from $f_0(\varepsilon_l)$ to $f_0(\tilde{\varepsilon}_l)$ must be through scattering.

The physical picture above can be justified by examining the Boltzmann equation, which describes the kinetics of electron wave packets. Assuming uniform system and field, it reads
$
    \partial_t f_{l}-\bm E \cdot \partial_{\bm k} {f_{l}}=\mathcal I \{f_{l}\}
$, where the right hand side is the collision integral, acting as a linear operator on $f_{l}$.
The $E$-field driving term here produces the usual non-equilibrium part $\propto E\tau$ in the distribution function, which, however, does not contribute to the intrinsic response.  Therefore, to study intrinsic response, one may drop this term and focus on the solution of
\begin{align}
    \partial_t f_{l}=\mathcal I \{f_{l}\}
    \label{eq}.
\end{align}

Note that this equation is describing wave packets with energies $\tilde{\varepsilon}_l$.
In other words, the scattering processes in $\mathcal I$ are defined for these dressed wave pacekts. For example,
for impurity scattering, we may write \cite{Luttinger1958}
\begin{equation}
  \mathcal I \{f_{l}\}=-\sum_{l'}(\omega_{l'l}f_l-\omega_{ll'}f_{l'}),
\end{equation}
with the scattering rate
$
  \omega_{l'l}=2\pi |T_{l'l}|^2 \delta (\tilde{\varepsilon}_{l'}-\tilde{\varepsilon}_l),
$
where $T$ is the scattering $T$-matrix \cite{sinitsyn_semiclassical_2007}.
Then, it is clear that $f_0(\tilde{\varepsilon}_l)$ is always a static solution of (\ref{eq}) (an explicit treatment of this simple case can be found in \cite{supp}). In contrast,
$f_0({\varepsilon}_l)$ is \emph{not} a static solution: If imposed as an initial distribution,
it should evolve by
Eq.~(\ref{eq}) and relax through scattering towards the static solution $f_0(\tilde{\varepsilon}_l)$.
In this process, their difference, i.e., IAD, comes about.

The above analysis is general, regardless of the details of the system as well as scattering. It follows that
there always exists the IAD $g_l$ in dc transport, which originates from scattering but its value is completely independent of scattering. Thus, in (D3), the intrinsic nonlinear current should be evaluated with $f_0(\tilde{\varepsilon}_l)$ instead of $f_0({\varepsilon}_l)$, and the result
indeed takes the form of Eq.~(\ref{jin}).

\textit{\color{blue} Frequency dependence.} It remains to find the result from (D2).
In this approach, we need to consider the frequency dependence under ac driving. Take $\tilde{E}_a(t)=E_a\cos\omega t=\text{Re}(E_a e^{i\omega t})$.
At $E^2$ order, the current response contains two components:
\begin{equation}
  j_a=\text{Re}(j_a^{(0)}+j_a^{(2\omega)}e^{i 2\omega t}),
\end{equation}
a zero-frequency (rectified) component $j^{(0)}$ and a second-harmonic component $j^{(2\omega)}$. Both may contain intrinsic contributions.

The distribution function can also be decomposed into frequency components, with
\begin{equation}
  f_l=f_0(\varepsilon_l)+\text{Re}\Big(g^{(0)}_l+g^{(\omega)}_l e^{i\omega t}+g^{(2\omega)}_l e^{i 2\omega t}\Big).
\end{equation}
It is known that  $g^{(\omega)}$ contributes only to extrinsic terms \cite{Lu2021}, so we focus on $g^{(0)}$ and $g^{(2\omega)}$.

From the analysis of Eq.~(\ref{eq}) just now, one can readily see that the rectified component $g^{(0)}$ always contains the following IAD
\begin{equation}\label{ADF rectified}
  g^{(0)}_l=f_0'(\varepsilon_l)\delta \varepsilon^{(0)}_l=-\frac{1}{4}f_0'(\varepsilon_l)G_{ab}^l E_a E_b,
\end{equation}
where $\delta \varepsilon^{(0)}_l$ denotes the zero-frequency component of the energy correction. This IAD is independent of frequency. Here, we assumed $\omega<\Delta$ with $\Delta$ the size of local gap at Fermi surface, such that the intraband transport dominates the response.

Meanwhile, for the second-harmonic component $g^{(2\omega)}$, its value depends on frequency and differs in two regimes. In the low-frequency regime $\omega\ll \tau^{-1}$, the left hand side of (\ref{eq}) is negligible, so the result takes the same form as the zero-frequency component:
\begin{equation}
  g_l^{(2\omega)}\approx f_0'(\varepsilon_l) \delta \varepsilon^{(2\omega)}_l=g^{(0)}_l.
\end{equation}
On the other hand, in the high-frequency regime $\tau^{-1}\ll\omega <\Delta$, the right hand side of (\ref{eq}) becomes negligible, and we obtain
\begin{equation}
  g_l^{(2\omega)}\approx 0.
\end{equation}
This can be readily understood, since at such high frequency, the oscillating electrons simply do not ``see'' the scatterers. For frequency between the two regimes, $g_l^{(2\omega)}$ should exhibit explicit $\omega$ dependence, and crossover from one regime to the other. In \cite{supp}, we show that under relaxation time approximation, the crossover takes the form of
\begin{equation}\label{ADF double}
  \text{Re}\Big(g_l^{(2\omega)}e^{i 2\omega t}\Big)=g^{(0)}_l\frac{\cos 2\omega t+2\omega\tau \sin 2\omega t}{1+(2\omega\tau)^2}.
\end{equation}

Inputting these results into Eq.~(\ref{current}), one finds that, in the sense of (D3), for the low-frequency regime
 $\omega\ll \tau^{-1}$, the ac intrinsic nonlinear current is given by
\begin{equation}\label{lowf}
  j^{(0)}_a=j^{(2\omega)}_a=\frac{1}{2}j^{\mathrm{in}}_a,
\end{equation}
where $j^{\mathrm{in}}_a$ is given by expression (\ref{jin}) with $E$ the \emph{amplitude} of ac $E$ field. On the other hand, for the high-frequency regime
$\tau^{-1}\ll\omega <\Delta$, the result is
\begin{equation}\label{highf}
  j^{(0)}_a=\frac{1}{2}j^{\mathrm{in}}_a,\qquad j^{(2\omega)}_a=\frac{1}{2}j^{\mathrm{cl}}_a.
\end{equation}
Therefore, the second-harmonic response exhibits difference between low and high frequency regimes,
whereas the rectified response does not. This is a unique feature of nonlinear response that has not been revealed before. Its experimental implications will be discussed in a while.

Now, we can find the result of dc intrinsic nonlinear current defined by (D2). This is by taking Eq.~(\ref{highf}) and extrapolating it to dc limit. The result is given by $(1/2)(\bm j^{\mathrm{cl}}+\bm j^{\mathrm{in}})$, which, in a sense, only partially captures IAD.

Therefore, the three approaches, which are considered equivalent for linear transport, give three different results for intrinsic nonlinear current.
All these results are summarized in Table~\ref{Difference}.

\textit{\color{blue} Which is the proper definition?} We have shown that the three approaches, which define the same intrinsic linear anomalous Hall response, produce distinct results when applied to nonlinear transport. The question is: Which one is appropriate? As clarified, their difference comes from IAD $g_l$, which has its root in scattering, but its contribution is completely independent of scattering. It arises from the natural local equilibration of electrons under a local shift of energy. It follows that in dc or low-frequency measurement, one can never separate $g_l$ from $f_0(\varepsilon_l)$; they always come together in the form of $f_0(\tilde{\varepsilon}_l)$ in producing the intrinsic response that is detected in experiment. IAD is an inseparable part of this response.
This shows (D3) is the \emph{proper} definition of intrinsic nonlinear current, since it corresponds to what is really measured in experiment.

The result from approach (D3), with $f_0(\tilde{\varepsilon}_l)$ for dc (low-frequency) transport, also has a clear physical picture and is physically appealing. In the picture of semiclassical theory, we are dealing with dynamics of a bunch of electron wave packets. The external fields not only supply driving force, they also dress the wave packets, leading to their renormalized energy $\tilde{\varepsilon}_l$ and Berry curvature $\tilde{\Omega}$. Then, in a consistent treatment, it is natural that the response is formulated in terms of these dressed
wave packets with their renormalized properties. In this sense, the use of $f_0(\tilde{\varepsilon}_l)$ for intrinsic response is a physically consistent choice. In the same spirit, for a consistent treatment, the scattering amplitude $|T_{l'l}|^2$ should also be defined with respect to the dressed wave packets, which has been shown in Ref.~\cite{Xiao2019NLHE}.

{\color{black}Our finding settles the controversy surrounding the experiments~\cite{Gao2023QM,Wang2023QM}. It indicates that the theoretical interpretation adopted in Ref.~\cite{Gao2023QM} is correct. The intrinsic nonlinear response probed in the experiment should correspond to Eq.~(\ref{jin}), which is purely Hall and forbidden by $C_{3z}$ symmetry, as verified in \cite{Gao2023QM}. Meanwhile, the interpretation [using Eq.~(\ref{jcl})] for experiment \cite{Wang2023QM} is incorrect.
The nonlinear signal that scales as $\sigma_{xx}^0$ found in that experiment should not be attributed to intrinsic response from MnBi$_2$Te$_4$ layers, because of the preserved $C_{3z}$. It is more likely to be attributed to so-called zeroth order extrinsic contributions~\cite{Huang2023scaling}, as supported by several recent studies~\cite{Huang2023scaling,Gao2024diode,Lu2024nonlinear}.
}

\textit{\color{blue} Discussion.} In this study, we have revealed the significant role of IAD in nonlinear transport.
It is worth noting that the nature of IAD is different from the so-called side-jump extrinsic contribution in anomalous Hall effect \cite{sinitsyn_semiclassical_2007}. In early literature \cite{nagaosa2010}, it is often stated that side-jump contribution is independent of scattering. However, this statement is invalid in general: side jump depends on the type of disorders and generally varies in the presence of more than one scattering sources \cite{Yang2011,hou2015}. In contrast, IAD is completely independent of these details, so it should be included in intrinsic response.

IAD arises from the local energy shift of electron wave packets. Implicit behind this picture is the separation between two time scales. One is the time scale for local equilibration among wave packets, which drives the distribution towards $f_0(\tilde{\varepsilon}_l)$. It is associated with the relaxation of the local energy shift $\delta\varepsilon_l$ (This is also the process that establishes $f_0(\varepsilon_l)$ in the absence of field). The other is the time scale for macroscopic transport, which might be regarded as a relaxation of electric potential energy. The latter is nonlocal in the sense that the relaxation requires the transport of electron across a macroscopic distance (through the sample). Clearly, in a diffusive system, the former time scale is much faster than the latter, which justifies the use of $f_0(\tilde{\varepsilon}_l)$ as a starting point for studying transport process.

We point out that the notion of IAD is \emph{universal}, and its importance is not limited to nonlinear transport effects.
It should be associated with any uniform energy shift for diffusive systems in steady state. {\color{black}Actually, we find that failing to incorporate IAD will lead to violation of fundamental physics relations or principles. For example,
it is known that on the basis of Einstein relation, the current response to $E$ field can be connected to that to chemical-potential gradient. The intrinsic nonlinear current driven by chemical-potential gradient was reported in Ref. \cite{gao2018}. In End Matter, we show that
it is Eq.~(\ref{jin}) from (D3), rather than Eq.~(\ref{jcl}), that fulfills the connection. In other words, IAD must be incorporated to be consistent with Einstein relation.} As a second example, it is well known that to linear order in a uniform magnetic field $B$, the electrons acquire a local energy shift $\delta\varepsilon_l=-\bm m_l\cdot\bm B$, where $\bm m_l$ is the spin plus orbital magnetic moment of the wave packet \cite{Zhang1996,xiao2010}. In the same spirit, this shift also induces an IAD, which must be considered. {\color{black}One can easily show that failing to account for this IAD would lead to a non-vanishing transport current driven a uniform dc $B$ field (see End Matter), which violates fundamental physics principles. }

Our discussion of IAD here is using the semiclassical theory framework, nevertheless, the revealed physics should be \emph{general}. This is because (1) it enjoys a transparent and coherent physical picture, and (2) the above evidences demonstrate it is indispensable for any theory that wants to maintain consistency with fundamental physical relations/principles.

{\color{black}We also mention that although phrased using familiar `classical' concepts, the semiclassical theory \emph{is} a quantum theory, which fully captures quantum effects (including interband coherence) in the intraband transport regime.
Meanwhile, how to adequately capture scattering effects for nonlinear transport using other quantum approaches is still a challenge. Rigorous treatment, e.g., the Kohn-Luttinger approach, becomes extremely complicated even for simple models.
To simplify, most recent works treat scattering by adding various phenomenological relaxation terms to quantum kinetic equation, which, however, produce unphysical and even self-contradictory results, as discussed in Ref. \cite{Jia2024}. In this regard, our work also offers an important guidance for theory development, namely, to achieve consistent physics, any proper treatment of scattering in a theory must at least capture the correct IAD contribution.}

Finally, as a by-product, we predict a new phenomenon for ac transport in the high-frequency regime $ \tau^{-1}\ll\omega< \Delta$. According to Eq.~(\ref{highf}) (and Table~\ref{Difference}), in this regime, due to the vanishing double-frequency IAD, the second-harmonic response $j^{(2\omega)}$ acquires a value different from the low-frequency regime and from the rectified component. This leads to a non-Hall and a longitudinal component of the second-harmonic nonlinear current. Using first-principles calculations, we perform an estimation of this phenomenon for antiferromagnetic CuMnAs (see End Matter) and predict a sizable response that can be detected in experiment.

%

\appendix
\renewcommand{\theequation}{A\arabic{equation}} 
\setcounter{equation}{0} 
\renewcommand{\thefigure}{A\arabic{figure}}
\setcounter{figure}{0}

\begin{center}
    \textbf{End Matter}
\end{center}

\emph{\color{blue}{Consistency with Einstein relation}}. {\color{black}According to the theory of electrodynamics, the \emph{transport} charge current in an inhomogeneous medium, e.g., due to chemical potential gradient $\bm\nabla\mu$ or temperature gradient $\bm\nabla T$,
is given by \cite{Jackson1999,Groot1972}
\begin{equation}
    \boldsymbol{j} = \boldsymbol{j}_\text{local} - \boldsymbol{\nabla}\times \boldsymbol{M} + \boldsymbol{\nabla}\times (\hat{e}_b \partial_a Q_{ab}).
\end{equation}
Here, $\boldsymbol{j}_\text{local}$ is the local current density, $\hat{e}_b$ is the unit vector along $b$, and the latter two terms are the circulating currents associated with magnetization $\boldsymbol{M}$ and magnetic quadrupole moment $Q_{ab}$. 

In Ref.~\cite{gao2018},
Gao and Xiao developed the theory of $Q_{ab}$ and obtained the following general formula for 
\emph{intrinsic} transport current accurate to the second order of statistical force ($\bm\nabla \mu$ or $\bm\nabla T$)~\footnote{In Ref.~\cite{gao2018}, the second term of the stated result contains a typo of sign and subscripts.}:
\begin{equation}\label{jG}
    \boldsymbol{j} = - \boldsymbol{\nabla} \times \sum_l \boldsymbol{\Omega}_l \mathcal{G}_l + \boldsymbol{\nabla} \times \Big( \hat{e}_b \partial_a \sum_l \theta_{ab}^l \mathcal{G}_l \Big),
\end{equation}
where 
\begin{equation}
  \mathcal{G}_l = -k_B T \ln\Big[1+\exp(\frac{\varepsilon_l-\mu}{k_BT})\Big]
\end{equation}
is the grand potential density, and $\theta_{ab}^l = \epsilon_{bcd}v_{c}G_{ad}^l$. The two terms in (\ref{jG}) are respectively the first-order and the second-order responses. 


Consider the transport current driven by $\boldsymbol{\nabla}\mu$. From the second term of (\ref{jG}), the intrinsic 
second-order nonlinear current can be obtained as
\begin{equation}\label{jmu}
    \boldsymbol{j}^{(2)} = - \sum_l f_0(\varepsilon_l)\Big[\boldsymbol{\nabla}\mu \times \left( \hat{e}_a \epsilon_{abc}\partial_{k_b}G_{cd}^l \partial_d \mu \right)\Big] .
\end{equation}

Now, by replacing $\bm\nabla\mu$ with $\bm E$, one immediately observes that this result exactly reproduces the intrinsic second-order current driven by $E$ field in Eq.~(\ref{jin}). This demonstrates the \emph{consistency} of Eqs.~(\ref{jin}) and (\ref{jmu}) with Einstein relation, which essentially states that $\bm\nabla\mu$ and $\bm E$ are equivalent in driving electric current.

IAD is essential in ensuring this consistency. It should be noted that IAD does not appear in the calculation of current driven by $\bm\nabla\mu$, as the statistical force $\bm\nabla\mu$ does not directly corrects $\varepsilon_l$ locally. Nevertheless, in studying such phenomena, it is well known that a prerequisite is the establishment of \emph{local} equilibrium in the system, so that $\mu(\bm r)$ [and $T(\bm r)$] are well defined. Therefore, the consistency with Eq.~(\ref{jin}) [from (D3)] found above is a clear manifestation that Eq.~(\ref{jin}), by capturing IAD, corresponds to the correct local equilibrium. In comparison, the other definitions (D1) and (D2) do not correctly capture IAD, thus violating the Einstein relation.

%
%
%
%
%

\emph{\color{blue}{Consistency with absence of transport driven by static $B$ field.}}
In condensed matter physics, it is a fundamental principle that a uniform static magnetic field alone cannot drive a transport current.


In the absence of $B$ field, the equilibrium Fermi distribution function is $f_0 (\varepsilon_{l})$, with the unperturbed band energy $\varepsilon_{l}$ as its argument. After applying a static $B$ field, the electron energy is changed to $\tilde{\varepsilon}_{l}=\varepsilon_{l}-\bm{m}_l\cdot \bm B$, where $\bm{m}_l$ is the magnetic moment of the electron wave-packet~\cite{Zhang1996,xiao2010}. This correction $-\bm{m}_l\cdot \bm B$ is also a local energy shift, just like $\delta \varepsilon_l$ in Eq.~(\ref{energy shift}). Thus, according to our analysis in the main text, the correct equilibrium distribution to use should be $f_0 (\tilde{\varepsilon}_{l})$, which incorporates the local equilibration process embodied by IAD.

Indeed, this treatment ensures the absence of electric current: 
\begin{equation}
  \bm j=\sum_l f_0(\tilde{\varepsilon}_l)\bm v_l=\sum_l f_0(\tilde{\varepsilon}_l)\partial_{\bm k}\tilde{\varepsilon}_l=\sum_l \partial_{\bm k}\mathcal{G}(\tilde{\varepsilon}_l)=0.
\end{equation}

On the other hand, if one misses the IAD contribution by still using distribution $f_0(\varepsilon_{l})$  to calculate the current, one would obtain
\begin{equation}
  \bm j=-\sum_l f_0(\varepsilon_{l})\partial_{\bm k} (\bm m_l\cdot\bm B),
\end{equation}
which is generally nonzero and hence violates the principle. This is a clear proof of the significance and the necessity of IAD.

The discussion can also be extended to time-varying (ac) $B$ field. Following our analysis in the main text regarding frequency dependence, for ac $B$ field in the low frequency regime ($\omega\ll \tau^{-1}$), the distribution should be 
$f_0 (\tilde{\varepsilon}_{l})$, which includes IAD. Meanwhile, for the high-frequency regime $\tau^{-1}\ll\omega<\Delta$, 
the electrons cannot `see' the scatterers, so IAD correction does not appear and the distribution is given by 
$f_0(\varepsilon_{l})$. The crossover between the two in the intermediate regime is described by~\cite{supp}:
\begin{equation}\label{fl}
 f_l= f_0 (\tilde{\varepsilon}_{l})+\text{Re}\Big(\frac{i\omega\tau }{1-i\omega\tau}g_l\Big),
\end{equation} 
where $g_l=f_0 (\tilde{\varepsilon}_{l})-f_0 ({\varepsilon}_{l})$ is the IAD. Therefore, a transport current can be driven by time-varying $B$ field in the high-frequency regime, but not in the low-frequency regime. This is consistent with many previous studies on chiral and gyrotropic magnetic effects~\cite{Pesin2015,Zhong2016}. Particularly, Eq.~(\ref{fl}) reproduces the result in Ref.~\cite{Zhong2016} obtained from Kubo linear response approach.
We stress that the understanding of IAD and its origin from scattering is a key to understand this frequency dependence. 
Without considering scattering at all, as in (D1), will cause incorrect results regarding frequency dependence~\cite{Chang2015PRB}.
}

\begin{figure}[t]
\begin{centering}
\includegraphics[width=8.5cm]{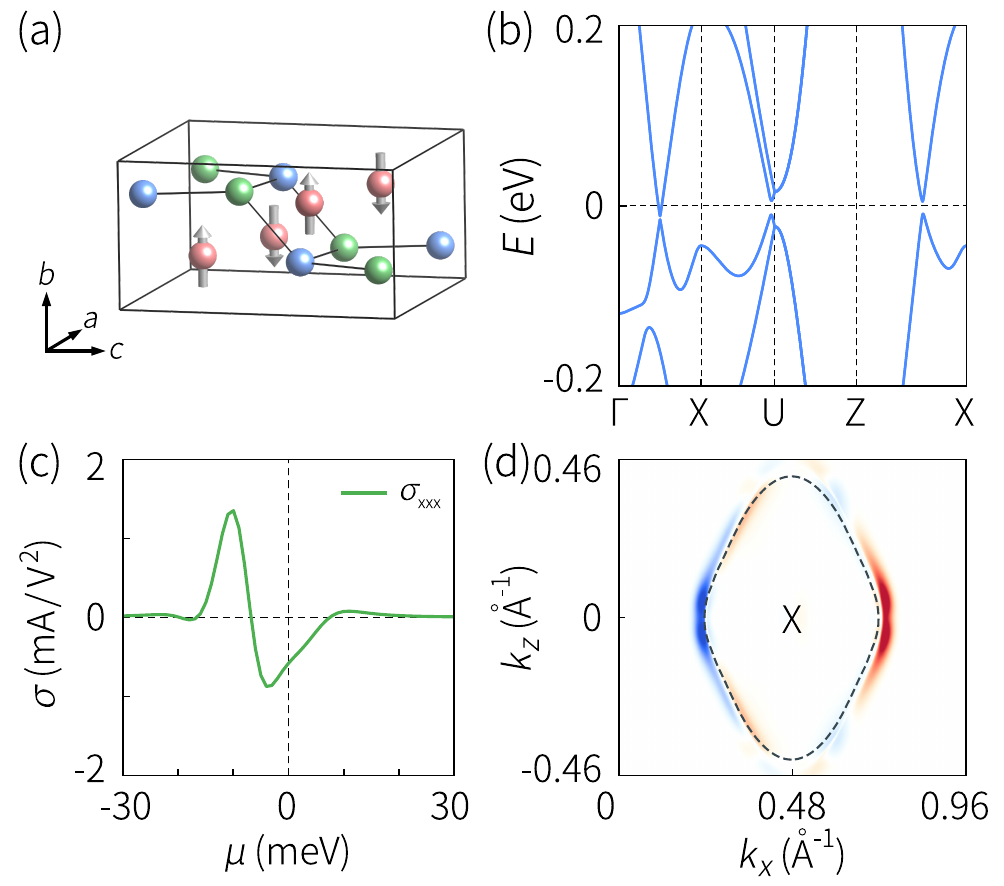}
\par\end{centering}
\caption{\label{Fig_CuMnAs}(a) Crystal structure of orthorhombic CuMnAs. The local moment direction in the antiferromagnetic ground state is marked. (b) Calculated band structure. (c) Calculated nonlinear response $\sigma^{(2\omega)}_{xxx}$ versus chemical potential $\mu$. Temperature is set at $30~$K. (d) Distribution of BCP dipole (the summand in Eq.~(\ref{long}) in the $k_y=0$ plane in momentum space. {The dashed line marks the nodal loop in the band structure around $X$ point, which is slightly gapped under spin-orbit coupling.}}
\end{figure}

\emph{\color{blue}{Estimation of nonlinear current response in CuMnAs in high-frequency regime.}}
According to Eqs.~(\ref{lowf}) and (\ref{highf}), one can see that at low frequencies $\omega\ll \tau^{-1}$, the $2\omega$ response is purely Hall, whereas in the high frequency regime
$ \tau^{-1}\ll\omega< \Delta$, then there will also appear a non-Hall component in the transverse current.
This can be distinguished in experiment by the sum-frequency generation method demonstrated in Ref.~\cite{Gao2023QM}. Meanwhile, in the high-frequency regime, there also emerges a second-harmonic component in longitudinal response. The corresponding
nonlinear conductivity is \cite{supp}
\begin{equation}\label{long}
    \sigma_{aaa}^{(2\omega)}=-\sum_l V_a^{ll} G_{aa}^l f'_{0}(\varepsilon_l)/4.
\end{equation}

To have an estimation of this response, we consider the example of orthorhombic CuMnAs, which is an antiferromagnet with N\'eel temperature $T_\mathrm{N}\approx300$~K~\cite{Muendelein1992Darstellung,Maca2012Room,Tang2016Dirac,Emmanouilidou2017Magnetic,Zhang2017Massive,ifmmodeSelseSfimejkal2017Electric,Huyen2021Spin}. Previous experiments~\cite{Emmanouilidou2017Magnetic,Zhang2017Massive} reported that its N\'eel vector is aligned along the $b$ axis [Fig.~\ref{Fig_CuMnAs}(a)]. The magnetic band structure obtained from first-principles calculations (details in \cite{supp}) is shown in Fig.~\ref{Fig_CuMnAs}(b), which agrees with previous calculations \cite{Emmanouilidou2017Magnetic}. This material belongs to the $m^{\prime}mm$ magnetic point group.
The symmetry-allowed nonlinear longitudinal response is $\sigma_{xxx}$. The result calculated from Eq.~(\ref{long}) is shown in Fig.~\ref{Fig_CuMnAs}(c). Interestingly, we find that the sizable $\sigma_{xxx}$ has a large contribution from
a nodal loop feature near the Fermi level [see Fig.~\ref{Fig_CuMnAs}(d)].
At $T=30~$K, the value of $\sigma_{xxx}$ can reach $\sim 1.2~$mA$/$V$^2$.  From previous experiment~\cite{Zhang2017Massive}, $\tau \sim 1$~ps at 30~K, and from the band structure, the local interband spacing $\Delta\sim 20$~meV ($30$~THz). Hence, we expect that our estimated scattering-independent nonlinear response $\sigma_{xxx}$ can be observed in the high-frequency window $\mathrm{0.5~THz} \ll \omega < \mathrm{30~THz}$.

\end{document}